\pdfoutput=1
\documentclass[12pt,preprint]{aastex}
\usepackage{graphicx}
\usepackage{natbib}
\usepackage{aas_macros}
\usepackage[section] {placeins}
\usepackage{subfigure}
\usepackage{float}
\usepackage{color}

\def\msun{{\rm\,M_\odot}}

\def\msun{{\rm\,M_\odot}}

\newcommand{\kms}{\, {\rm km\, s}^{-1}}

\newcommand{\mpc}{\, {\rm Mpc}}
\newcommand{\kpc}{\, {\rm kpc}}
\newcommand{\hmpc}{\, h^{-1} \mpc}

\newcommand{\hi}{\mbox{H\,{\scriptsize I}\ }}
\newcommand{\heii}{\mbox{He\,{\scriptsize II}\ }}

\def\h2{${\rm\,H_2}$}

\makeatletter 

\def\hi{\noindent \hangindent=2.5em}

\def\kpc{{\rm\,kpc}}

\def\mpc{{\rm\,Mpc}}

\def\kms{{\rm\,km/s}}
\def\msun{{\rm\,M_\odot}}

\def\vol#1  {{{#1}{\rm,}\ }}

\def\hi{{H\thinspace I}\ }

\def\heii{{He\thinspace II}\ }

\newcount\refno
\newcount\rfno
\def\eq{$^{\the\refno\ }$\advance\refno by 1}
\def\ad{\advance\rfno by 1}

\def\clock{\count0=\time \divide\count0 by 60
     \count1=\count0 \multiply\count1 by -60 \advance\count1 by \time
     \number\count0:\ifnum\count1<10{0\number\count1}\else\number\count1\fi}

\def\myputfigure#1#2#3#4#5%
{\hskip0.03\textwidth\vskip#5pt
\makebox[0pt]{\hskip#2in
\includegraphics[width=#3\textwidth]{#1}}\vskip#4pt\hfill}

\newcount\refno
\newcount\rfno
\def\eq{$^{\the\refno\ }$\advance\refno by 1}
\def\ad{\advance\rfno by 1}

\definecolor{burntorange}{rgb}{1,0.4,0.2}

\begin{document}

\title{Frequent Spin Reorientation of Galaxies due to Local Interactions}

\author{
Renyue Cen$^{1}$
} 

\footnotetext[1]{Princeton University Observatory, Princeton, NJ 08544;
 cen@astro.princeton.edu}

\begin{abstract} 

We study the evolution of angular momenta of ($M_*=10^{10}-10^{12}\msun$) galaxies
utilizing large-scale ultra-high resolution cosmological hydrodynamic simulations
and find that spin of the stellar component changes direction frequently,
caused by major mergers, minor mergers, significant gas inflows and torques by nearby systems. 
The rate and nature of change of spin direction can not be accounted for 
by large-scale tidal torques, because the latter fall short in rates by orders of magnitude
and because the apparent random swings of the spin direction are inconsistent with alignment by linear density field. 
The implications for galaxy formation as well as intrinsic alignment of galaxies are profound. 
Assuming the large-scale tidal field is the sole alignment agent, a new picture emerging is that intrinsic alignment of galaxies 
would be a balance between slow large-scale coherent torquing and fast spin reorientation by local interactions.
What is still open is whether other processes, such as feeding galaxies with gas and stars along filaments or sheets,
introduce coherence for spin directions of galaxies along the respective structures.

\end{abstract}


\section{Introduction}

The angular momentum or spin of galaxies is a physical quantity that is far from being fully understood but is 
of fundamental importance to galaxy formation and cosmological applications.
While N-body simulations have shed useful light on spin properties of dark matter halos \citep[e.g.,][]{2002Vitvitska},
it is expected that, given the vastly different scales between the stellar component and dark matter halo component and different physical processes
governing stellar, gas and dark matter components, 
the angular momentum dynamics of galaxies may be quite different and not necessarily inferable from N-body simulations with any reasonable accuracy. 
We herewith perform a detailed analysis of the dynamics of spin of galaxies in a full cosmological context,
utilizing {\it ab initio} LAOZI cosmological hydrodynamic simulations
of the standard cold dark matter model \citep[][]{2014Cen} with an unprecedented galaxy sample size and ultra-high numerical resolution.
This paper is the second in the series ``On the Origin of the Hubble Sequence".

\section{Method}\label{sec: sims}

The reader is referred to \citet[][]{2014Cen} for detailed descriptions 
of our simulations and validations.
Briefly, we perform cosmological simulations with the adaptive mesh refinement hydrocode, Enzo \citep[][]{2013Enzo}.  
The periodic box has a size of $120h^{-1}$Mpc,
within which a zoom-in box of a comoving size of $21\times 24\times 20h^{-3}$Mpc$^3$ is emdedded.
The resolution is better than $114h^{-1}$pc (physical).
The cosmological parameters are the same as the WMAP7-normalized \citep[][]{2010Komatsu} $\Lambda$CDM model.

We identify galaxies using the HOP algorithm 
\citep[][]{1998Eisenstein} operating on the stellar particles. 
A sample of $\ge 300$ galaxies with stellar masses greater than $10^{10}\msun$ are used. 
For each galaxy at $z=0.62$ a genealogical line is constructed from
$z=0.62$ to $z=6$ by connecting galaxy catalogs at a series of redshifts.
Galaxy catalogs are constructed from $z=0.62$ to $z=1.40$ at a redshift increment of $\Delta z=0.02$ 
(corresponding to $\Delta t=81$Myr at $z=1$) and 
from $z=1.40$ to $z=6$ at a redshift increment of $\Delta z=0.05$
(corresponding to $\Delta t=80$Myr at $z=2$).
The parent of each galaxy is identified with the one at the next higher redshift catalog that has the most overlap in stellar mass.

We compute the specific angular momentum vector ${\vec{\bf j}}_{i}$ for stars
of each galaxy within a radius $r$ 
at each output snapshot $i$.
The time derivative of ${\vec {\bf j}}_i$ is computed as 
\begin{equation}
\label{eq:djdt}
|d{\vec{\bf j}}_i/dt|\equiv |{\vec{\bf j}}_{i+1}-{\vec{\bf j}}_{i}|/(t_{i+1}-t_{i}).
\end{equation}
\noindent
One notes that due to the finite number of outputs for our simulation data,
$d{\vec {\bf j}}_*/dt$ is somewhat underestimated in cases of rapid changes of angular momentum on time scales shorter than our snapshot intervals.
A similar definition for gas is also used.
We denote $t_{1}$ as the time required to change the spin vector by $1$ degree of arc at each snapshot for each galaxy,
defined as 
\begin{equation}
\label{eq:t1}
t_{1}\equiv {\pi\over 180} (t_{i+1}-t_{i}){\rm acos}^{-1}({\hat{\bf j}}_{i+1}\cdot \hat{\bf j_{i}}),
\end{equation}
\noindent
where ${\hat{\bf j}}_{i}$ is the unit vector of ${\vec{\bf j}}_{i}$.
For the first time, we address the evolution of the spin of galaxies statistically in a cosmological setting.
All length units below will be physical.

\section{Results}

\begin{figure}[ht!]
\centering
\vskip -0.0cm
\resizebox{6.0in}{!}{\includegraphics[angle=0]{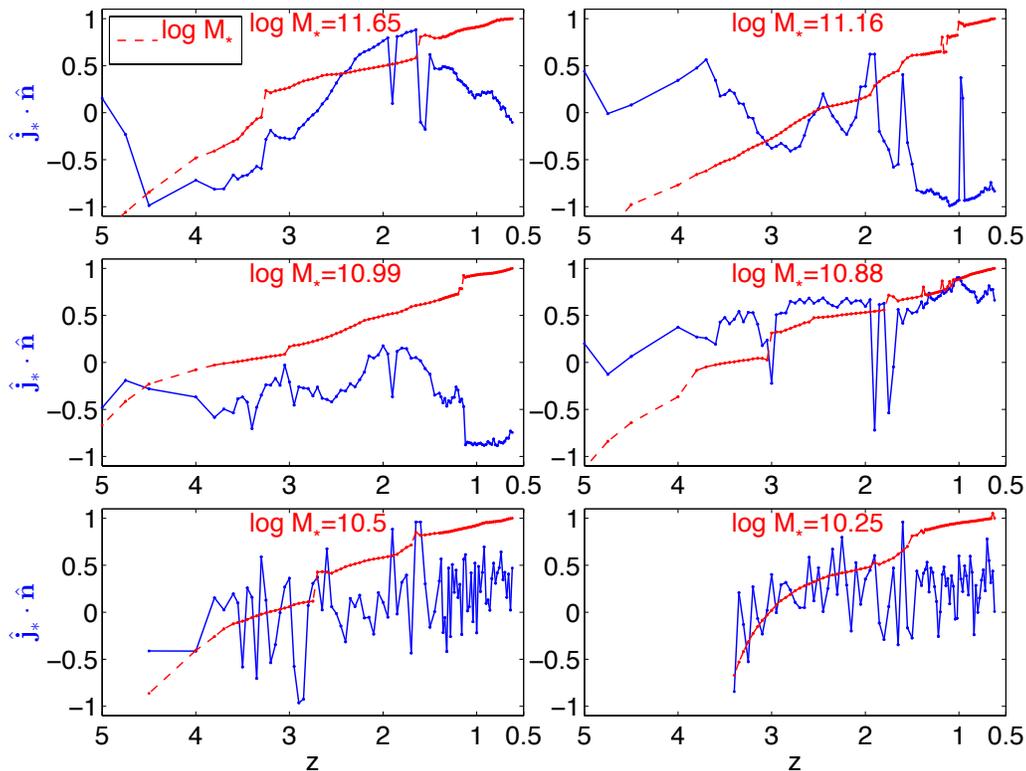}}   
\vskip -0.5cm
\caption{
shows in blue the dot product of the unit vector of the specific angular momentum of the central 3kpc 
stellar region and an arbitrary fixed (in time) unit vector 
as a function of redshift.
Each panel shows a random galaxy with its final stellar mass at $z=0.62$ as indicated at the top of the panel.
Also shown in each panel as a red dashed line is the logarithm of the stellar mass with an arbitrary vertical offset.
}
\label{fig:J3northz}
\end{figure}

Figure~\ref{fig:J3northz} shows 
the dot product of the unit vector of the specific angular momentum of the central 3kpc radius
stellar region and an arbitrary fixed unit vector 
as a function of redshift in blue.
It is visible that a significant increase in stellar mass within a short period of time
(i.e., mergers) is often accompanied by dramatic changes in angular momentum vectors. 
We note that the angular momentum vector of a galaxy over its history 
displays a substantial amount of change even in ``quiet" times without major mergers.
Ensuing analysis provides some physical insight into this.

\begin{figure}[ht!]
\centering
\vskip -0.0cm
\resizebox{6.0in}{!}{\includegraphics[angle=0]{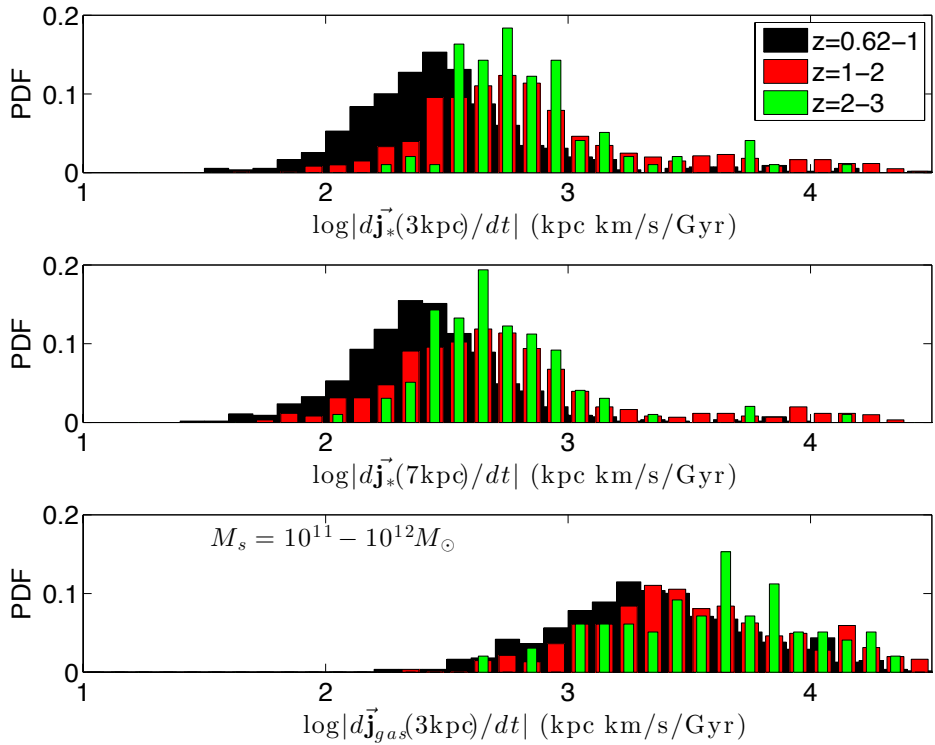}}   
\vskip -0.5cm
\caption{
{\color{burntorange}\bf Top panel:} 
the probability distribution function (PDF) of the amplitude of the time derivative of specific angular momentum of 
the central $3$kpc radius
stellar regions (see Eq ~\ref{eq:djdt})
for galaxies of total stellar mass in the range $10^{11}-10^{12}\msun$
in three different redshift ranges, $z=0.62-1$ (black histograms), $z=1-2$ (red histograms), $z=2-3$ (green histograms),
respectively.
As an intuitive example, if a Milky Way-like galaxy of size $10$kpc and rotation velocity of $200$km/s
changes its spin direction by 90 degress in one current Hubble time,
it would correspond to a value $\log|dj/dt|$ equal to $2.3$ in the x-axis.
{\color{burntorange}\bf Middle panel:} 
same as the top panel but for the central $7$kpc radius stellar region.
{\color{burntorange}\bf Bottom panel:} 
same as the top panel but for gas in the central $3$kpc radius region. 
}
\label{fig:dJdt}
\end{figure}

The top panel of Figure~\ref{fig:dJdt} shows 
the PDF of the time derivative of specific angular momentum of 
the central $3$kpc radius stellar regions for galaxies of stellar mass in the range $10^{11}-10^{12}\msun$.
The middle panel shows the same as in the top panel, except it is for the central $7$kpc radius stellar regions. 
We see that the overall rate of change of angular momenta is significantly higher at $z=1-3$ compared to that at $z=0.6-1$.
The distribution of $|d{\vec{\bf j}}_*/dt|$ has an extended tail at the high-end, due to major mergers;
due to our finite time sampling these rates are capped by the frequency of our snapshots.
Consistent with the expected decline of major merger rate below $z\sim 1$,
the high $|d{\vec{\bf j}}_*/dt|$ tail of the distribution at $z=0.6-1$ is significantly less pronounced.
No major difference is seen between $3$kpc and $7$kpc cases, 
suggesting that angular momentum changes within the two radii are approximately 
in tandem and our analysis is robust using $3$kpc.
The choice of $3-7$ proper kpc is appropriate by noting that a (spiral, elliptical) galaxy of stellar mass $10^{12}\msun$
is observed to have a size of $(10.8, 15.1)$kpc \citep[][]{2003Shen} for low redshift galaxies.
The size roughly scales with the root of the stellar mass and decreases with 
increasing redshift \citep[e.g.,][]{2006Trujillo}.

\begin{figure}[!h]
\centering
\vskip -0.5cm
\hskip -1.0cm
\resizebox{7.25in}{!}{\includegraphics[angle=0]{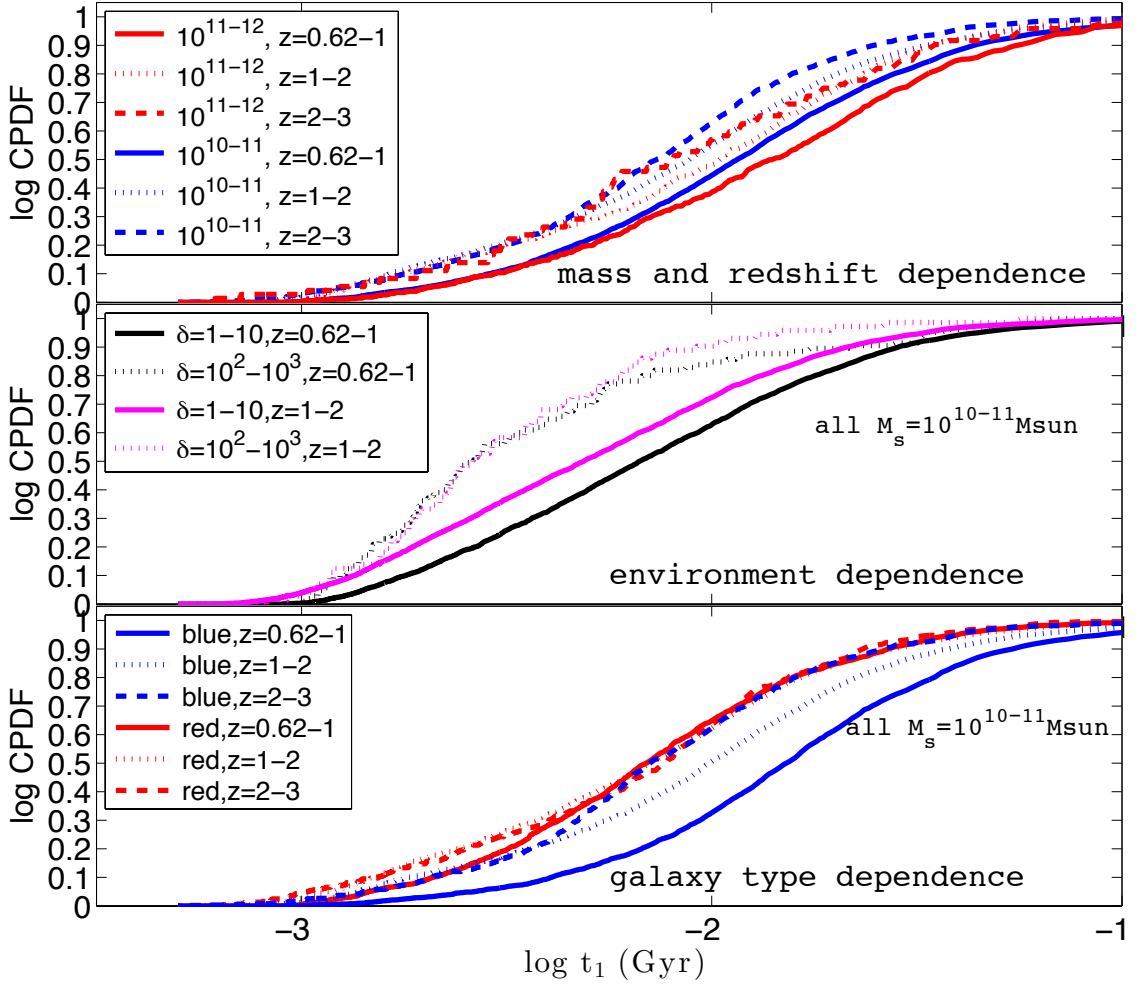}}   
\vskip -0.5cm
\caption{
{\color{burntorange}\bf Top panel:} shows the cumulative PDF (CPDF) of the time taken
to change the direction of spin of the central $3$kpc radius
stellar region by $1$ degree of arc, $t_1$ (Eq ~\ref{eq:t1})
for galaxies of total stellar mass in the range $10^{10}-10^{11}\msun$ (blue curves) and $10^{11}-10^{12}\msun$ (red curves)
in three different redshift ranges, $z=0.62-1$ (solid curves), $z=1-2$ (dotted curves), $z=2-3$ (dashed curves), respectively.
{\color{burntorange}\bf Middle panel:} shows the of $t_1$ for galaxies of total stellar mass in the range $10^{10}-10^{11}\msun$
in low-density ($\delta_{0.5}=1-10$); solid curves) and high-density environment ($\delta_{0.5}=10^2-10^3$); dotted curves), 
in two redshift rangess, $z=0.62-1$ (black curves) and $z=1-2$ (magenta curves), respectively.
The environment overdensity $\delta_{0.5}$ is defined to be the overdensity of total matter in a sphere of radius $0.5h^{-1}$Mpc comoving.
{\color{burntorange}\bf Bottom panel:} shows the CPDF of $t_1$
for blue ($g-r<0.6$; blue curves) and red ($g-r>0.6$, red curves) galaxies of total stellar mass in the range $10^{10}-10^{11}\msun$ 
in three different redshift ranges, $z=0.62-1$ (solid curves), $z=1-2$ (dotted curves), $z=2-3$ (dashed curves), respectively.
}
\label{fig:cosdt}
\end{figure}

The bottom panel of Figure~\ref{fig:dJdt} 
shows the PDF for gas in the central $3$kpc 
radius
region.
We see that the specific angular momenta of the gas within central $3$kpc 
change at rates $5-10$ times higher than that of stars (top panel of Figure~\ref{fig:dJdt}).  
There is no doubt that 
gas inflows contribute significantly to the change of the stellar angular momentum in two ways.
First, significant gas inflows at inclined angles to the stellar mid-plane may torque the stars (and vice versa).
Second, new gas that reaches there 
will form new stars that have different angular momentum vector and cause the overall angular momentum to change in both direction and magnitude.
At high redshift the orientation of the gas inflows on large scales are not well correlated with that of the stars or gas that
is already there. Since the amount of gas tend to be smaller than that
of stars, it is easier to alter the angular momentum of the gas than that of the stars. 
In the absence of major mergers, we expect minor stellar mergers could also alter the angular momentum vector.

Figure~\ref{fig:cosdt} shows the CPDF of the time to change the direction of spin of the central $3$kpc radius stellar region by $1$ degree of arc,
for dependence on mass and redshift (top panel), environment (middle panel) and galaxy type (bottom panel).
Consistent with Figure~\ref{fig:dJdt} we see that 
the frequency of spin direction change increases with redshift;
the median $t_1$ decreases by $60-80\%$ 
from $z=0.62-1$ to $z=2-3$ with the higher mass group corresponding to the high end of the range of change.
The median $t_1$ decreases by $10-20\%$ from 
$M_{s}=10^{11-12}\msun$ to $M_{s}=10^{10-11}\msun$
with the dependence on mass somewhat stronger at low redshift than at high redshift. 
That less massive galaxies tend to experience more rapid changes of specific angular momenta is anecdotally apparent in Figure~\ref{fig:cosdt}.
We also find large mis-alignment between inner stellar (and gas) regions with outer halos (not presented here),
in broad agreement with the conclusions of \citet[][]{2010Hahn}. 

A dependence on environment is seen in the middle panel,
with the median $t_1$ decreasing by a factor of $1.9-2.7$ from $\delta_{0.5}=1-10$ to $\delta_{0.5}=10^2-10^3$; 
the environment dependence weakens at higher redshift.
This finding that the spin direction of galaxies changes more frequently in dense environment 
can be attributed to enhanced local interactions there.
In the bottom panel the dependence on galaxy type gives mixed trends.
For blue ($g-r<0.60$) galaxies the median $t_1$ decreases steadily from $z=2-3$ to $z=0.62-1$ by a factor of $\sim 2.3$,
whereas for red ($g-r>0.60$) galaxies the median $t_1$ hardly changes from $z=2-3$ to $z=0.62-1$.
The median $t_1$ for red galaxies is comparable to that of blue galaxies at $z=2-3$;
at lower redshift the median $t_1$ for red galaxies becomes progressively lower compared to that
of blue galaxies, mainly due to the latter increasing with decreasing redshift.
In \citet[][]{2014Cen} we show that the vast majority of red galaxies do not gain significant stellar mass in the red sequence.
Thus we conclude that the rapid change of spin direction for red galaxies
are due to torques by nearby galaxies, whereas blue galaxies are subject to all three local interactions
- gas accretion, stellar accretion and torques.

It is instructive to put the frequency of spin direction change into some perspective.
For a point mass of $M$ at a distance $d$, 
the torque of $M$ on the galaxy with a quadrupole moment $Q$ and angular momentum $\vec J$ is
$\tau =|{d\vec J\over dt}| = {3\over 4} {G M Q\over d^3}\sin (2\theta)$ \citep[e.g.,][]{1969Peebles}, 
where $\theta$ is the angle between the separation vector and the symmetry axis of the galaxy.
Expressing $\tau$ in terms of overdensity $\delta$ of the region centered on mass $M$,
$\tau = \pi G \rho_0 (1+z)^3 \delta Q\sin (2\theta)$, where $z$ is redshift, $\rho_0$ the mean mass density at $z=0$.
Approximating spirals as flat axisymmetric uniform disks with $a=b=\infty c$
(giving the quadrupole moment of $Q=2ma^2/5$) and full rotation support.
This allows to express the torquing time $t_{\rm q}$, defined to be the time taken to change the spin direction by $1$ degree of arc, 
\begin{equation}
\label{eq:tq}
t_{\rm q} = {\pi\over 180} {|\vec {\bf j}_*|m\over\tau(z,m,T)},
\end{equation}
giving $t_{\rm q}=2.5$Gyr for $\delta=200$, $z=1$ and $\sin(2\theta)=1$ for spiral galaxies. 
Comparing to the median $t_1\sim 10^{-3}-10^{-2}$Gyr seen in Figure~\ref{fig:cosdt},
it is evident that the rapid spin reorientation of galaxies can not possibly be due to tidal torques by large-scale structure.
It is noted that the intrinsic alignment sourced by primordial large-scale gravitational field is 
inconsistent with the frequent directional change shown in Figure~\ref{fig:J3northz}. 

Under the (unproven) assumption that the large-scale tidal field is the sole alignment agent,
any alignment between galaxies on large-scales would result from a balance between the fast reorientation rate due to 
local processes and slow coherent torques by large-scale structure,  
which is expressed as the ratio $t_1$ to $t_{\rm q}$, denoted as $t_1/(t_1+t_{\rm q})$.
If the quadrupole of the galaxy is, in this case, produced by loal interactions, independent of the large-scale tidal field,
the alignment in this simplified model would be linear [instead of quadratic, see \citet[][]{2004Hirata}] to the large-scale gravitational tidal field.
We obtain finally the expression for the mean value of $t_1/(t_1+t_{\rm q})$ 
weighted by the distribution of $t_1$ [$P(t_1)$, shown in Figure ~\ref{fig:cosdt}], denoted as $\eta(z,m,T)$:
\begin{equation}
\label{eq:eta}
\eta(z,m,T) \equiv \int \left({t_1\over t_1+t_{\rm q}(z,m,T)}\right)P_{z,m,T}(t_1)dt_1,
\end{equation}
at redshift $z$ for galaxies of mass $m$ and type $T$ (spiral or elliptical).
We approximate elliptical galaxies as oblate axisymmetric spheroids with $a=b=2c$ and $v_{\rm rot}/\sigma=0.2$,
resulting in the quadrupole moment of $Q=3ma^2/10$.
The sizes of galaxies are adopted from observations by \citet[][]{2003Shen}.
The bias factor is from \citet[][]{2004Tegmark} adjusted to $\sigma_8=0.8$.
The stellar mass to light ratio as a function of absolute magnitude is taken from from \citet[][]{2003bKauffmann}. 
We incorporate these into $\tau$ to get 
\begin{equation}
\label{eq:tau}
\tau(z,m,T) = \pi G \rho_0 (1+z)^3 D_+(z) b(m) Q(m,T) [\delta\sin(2\theta)]
\end{equation}
\noindent
as well as into $j$ in Equations (\ref{eq:tq}) for different galaxy types,
where $D_+(z)$ is the linear density growth factor normalized to be unity at $z=0$,
$b(m)$ is the bias factor of galaxies of mass $m$,

\begin{figure}[ht!]
\centering
\vskip -0.5cm
\resizebox{6.0in}{!}{\includegraphics[angle=0]{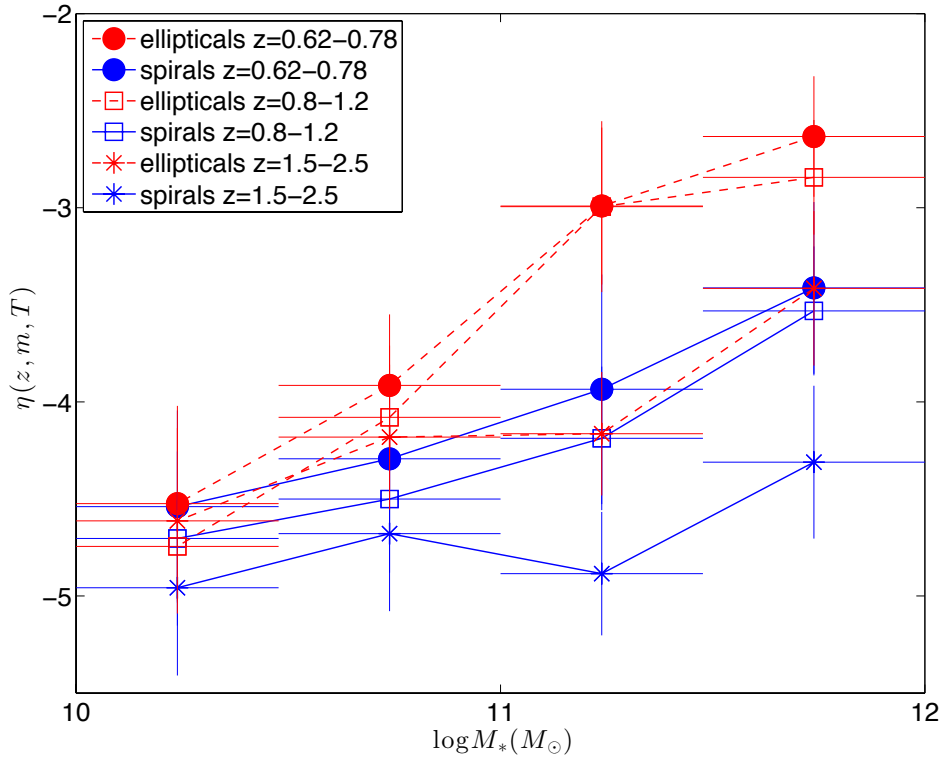}}   
\vskip -0.5cm
\caption{
shows $\eta$ (see Equation \ref{eq:eta} for definition) as a function of galaxy mass and type at three different redshift ranges.
Spiral and elliptical galaxies are shown are shown in blue and red, respectively,
for $z=0.62-0.78$ (solid dots), $z=0.8-1.2$ (open squares) and $z=1.5-2.5$ (stars). 
The model results are obtained using $[\delta\sin(2\theta)]=1$ (see Equation \ref{eq:tau}).
The errorbars in the x-axis indicate the mass bin sizes.
}
\label{fig:align}
\end{figure}

The results using Equation (\ref{eq:eta}), in conjunction with Equations (\ref{eq:tq}, \ref{eq:tau}),
are shown in Figure~\ref{fig:align}.
As in the linear alignment model \citep[e.g.,][]{2001Catelan}, 
the difficulty is to define a demarcating scale between local and linear large-scale structures.
We tentatively have left the scalings to be relative, absorbed into $[\delta\sin(2\theta)]$.
If compelled to give an estimate relevant to weak lensing,
one might choose $[\delta\sin(2\theta)]$ to be in the range $1-10$.
In this case, we get a tangential sheer $\gamma_{\rm T}$
that is $1-10$ times $\eta$ in Figure~\ref{fig:align}, resulting in $\gamma_{\rm T}$
of $-(0.2-2)\%$ for the most massive elliptical galaxies (i.e., luminous red galaxies, LRGs, red dots in Figure~\ref{fig:align}), 
which, coincidentally,
falls in the range of observed GI (galaxy-gravitational tidal field) signal for LRGs \citep[e.g.,][]{2006Mandelbaum, 2007Hirata, 2011Joachimi}.
The negative sign comes about, because the galaxies, under the torque of a central mass,
have a tendency to align their disks in the radial direction that is dynamically stable.

Three separate trends with respect to $z$, $m$ and $T$ are seen:
the alignment (1) decreases with increasing redshift,
(2) decreases with decreasing stellar mass,
and (3) is larger for elliptical galaxies than for spiral galaxies.
The first two trends are accounted for by trends of $t_1$ seen in Figure ~\ref{fig:cosdt}.
The last trend requires some discussion.
The bottom panel of Figure ~\ref{fig:cosdt} shows that
ellipticals have shorter $t_1$ than spirals, due to a large part to their
residing in overdense environments 
and in addition to their having a lower overall specific angular momentum amplitude.
However, also because the specific angular momentum of ellipticals is a factor of $5$ 
lower than that of spiral galaxies, elliptical galaxies are easier to slew.

It is argubly a relatively more straight-forward comparison
to observations of (radial) alignments of satellite galaxies with respect to the central galaxies of groups and clusters.
But this is in fact complicated by (at least) four issues.
First, the observed detection and non-detection of radial alignment of galaxies
around groups and clusters of galaxies concern radial ranges that are already mostly in the nonlinear regime (i.e., overdensity $\delta \gg 1$).
Second, most of the observed galaxy samples analyzed contain of order 100-10000 galaxies, hence statistical uncertainties are in the range of $1-10\%$.
Third, observed samples likely contain a large number of projected galaxies with physical separations
that are much larger than their lateral distance from the cluster/group center;
the degree of projection effects is strongly dependent on the orientation of the line of sight
(e.g., viewing a cluster along a filament) and significantly complicates interpretation of results.
Fourth, on some very small scales, binary interactions between a satellite and the central galaxy may play the dominant role.
A combination of these factors may explain the current confused state with conflicting observational results
\citep[e.g.,][]{2002Bernstein, 2005Pereira, 2006Agustsson, 2007Torlina, 2007Faltenbacher, 2011Hao}.
Nonetheless, we expect that the radial alignment, if exists, is expected to decrease
with increasing redshift, perhaps already hinted by some observations \citep[e.g.,][]{2012Hung},
and with decreasing cluster mass at a fixed radius.

The simple model presented has two notable caveats.
First, it assumes that
the only alignment mechanism is gravitational torque by some large-scale 
structure. So far we have presented only the relative scalings among different galaxies
under this assumption, but not the absolute magnitude.
We cannot justify this rather critical assumption with confidence at this time.
Second, one notes that a significant portion of the galaxy spin direction reorientation  
is likely due to gas feeding and substructure merging.
Thus, it is not unreasonable to expect that the gas feeding and substructure merging
have some preferred directions, such as along the filaments and sheets.
In this case, while galaxy spin direction changes frequenctly as shown here, 
it may do so with some degree of coherence over some scales (such as the scale of filaments), 
either temporaneously or through long-term memory of large-scale structure \citep[e.g.,][]{2012Libeskind}.
If this were true, it then suggests that intrinsic alignments may 
be a result of balance between high-frequency random re-orientation at short time scales 
and some sort of large-scale ``mean" feeding pattern on long time scales.
There is some empirical evidence for galaxies to be aligned with large-scale structures in a sense that 
is consistent with this ``feeding" picture \citep[e.g.,][]{2013Zhang,2013Li}.
It should be a priority to understand this issue systematically.

\section{Conclusions}

Utilizing {\it ab initio}
{\color{red}\bf L}arge-scale {\color{red}\bf A}daptive-mesh-refinement 
{\color{red}\bf O}mniscient {\color{red}\bf Z}oom-{\color{red}\bf I}n cosmological hydrodynamic simulations 
({\color{red}\bf LAOZI Simulation}) of the standard cold dark matter model,
we study the evolution of angular momenta of massive ($M_*=10^{10}-10^{12}\msun$) galaxies.
The simulation has an ultra-high resolution of $\le 114$pc/h and contains more than 300 galaxies
with stellar mass greater than $10^{10}\msun$. 
We find that spin of the stellar component changes direction frequently,
caused by major mergers, minor mergers, significant gas inflows and torques by nearby systems, 
with a median time in the range $1-10$Myr for directional change of spin vector by $1$ degree of arc.
The spin of the gas component changes at a factor of $5-10$ higher rates than the stellar component.
Because the processes that are responsible are mostly in the nonlinear regime.
we do not expect that the findings significantly depend on precise cosmological parameters,

The rate of change of spin direction can not be accounted for 
by large-scale tidal torques, because the latter fall short in rates by $2-3$ orders of magnitude.
In addition, the nature of change of spin direction - apparent random swings -
is inconsistent with alignment by linear density field. 
A new paradigm emerging with respect to intrinsic alignment of galaxies 
is that it is determined, primarily, by a balance between slow large-scale coherent torquing 
(if it were the sole alignment process) and fast spin reorientation by local interactions.
This suggests that a significant revision to the large-scale tidal torque based alignment theory is perhaps in order.
A simple analysis presented here indicates that intrinsic alignment of galaxies is dependent on redshift, luminosity, environment and galaxy type.
Specifically, it is found that the alignment (1) decreases with increasing redshift,
(2) decreases with decreasing stellar mass, and (3) is larger for elliptical galaxies than for spiral galaxies.
While no detailed comparisons are made, the found trends appear to be 
broadly consistent with and thus provide the physical basis for the observed trends.

What remains open is whether other processes, such as feeding galaxies with gas and stars along filaments or sheets,
introduce some coherence of their own kind for spin direction of galaxies along the respective structures.
This will require a separate study in greater detail.

\vskip 1cm

I would like to thank Claire Lackner for providing the SQL based merger tree construction software,
and \citep[][]{2011Turk} for providing the very useful analysis and visualization program yt.
Computing resources were in part provided by the NASA High-End Computing (HEC) Program through the NASA Advanced
Supercomputing (NAS) Division at Ames Research Center.
This work is supported in part by grant NASA NNX11AI23G.
The simulation data are available from the author upon request.


\begin{thebibliography}{27}
\expandafter\ifx\csname natexlab\endcsname\relax\def\natexlab#1{#1}\fi

\bibitem[{{Agustsson} \& {Brainerd}(2006)}]{2006Agustsson}
{Agustsson}, I., \& {Brainerd}, T.~G. 2006, \apjl, 644, L25

\bibitem[{{Bernstein} \& {Norberg}(2002)}]{2002Bernstein}
{Bernstein}, G.~M., \& {Norberg}, P. 2002, \aj, 124, 733

\bibitem[{{Catelan} {et~al.}(2001){Catelan}, {Kamionkowski}, \&
  {Blandford}}]{2001Catelan}
{Catelan}, P., {Kamionkowski}, M., \& {Blandford}, R.~D. 2001, \mnras, 320, L7

\bibitem[{{Cen}(2014)}]{2014Cen}
{Cen}, R. 2014, \apj, 781, 38

\bibitem[{Eisenstein \& Hut(1998)}]{1998Eisenstein}
Eisenstein, D.~J., \& Hut, P. 1998, ApJ, 498, 137

\bibitem[{{Faltenbacher} {et~al.}(2007){Faltenbacher}, {Li}, {Mao}, {van den
  Bosch}, {Yang}, {Jing}, {Pasquali}, \& {Mo}}]{2007Faltenbacher}
{Faltenbacher}, A., {Li}, C., {Mao}, S., {van den Bosch}, F.~C., {Yang}, X.,
  {Jing}, Y.~P., {Pasquali}, A., \& {Mo}, H.~J. 2007, \apjl, 662, L71

\bibitem[{{Hahn} {et~al.}(2010){Hahn}, {Teyssier}, \& {Carollo}}]{2010Hahn}
{Hahn}, O., {Teyssier}, R., \& {Carollo}, C.~M. 2010, \mnras, 405, 274

\bibitem[{{Hao} {et~al.}(2011){Hao}, {Kubo}, {Feldmann}, {Annis}, {Johnston},
  {Lin}, \& {McKay}}]{2011Hao}
{Hao}, J., {Kubo}, J.~M., {Feldmann}, R., {Annis}, J., {Johnston}, D.~E.,
  {Lin}, H., \& {McKay}, T.~A. 2011, \apj, 740, 39

\bibitem[{{Hirata} {et~al.}(2007){Hirata}, {Mandelbaum}, {Ishak}, {Seljak},
  {Nichol}, {Pimbblet}, {Ross}, \& {Wake}}]{2007Hirata}
{Hirata}, C.~M., {Mandelbaum}, R., {Ishak}, M., {Seljak}, U., {Nichol}, R.,
  {Pimbblet}, K.~A., {Ross}, N.~P., \& {Wake}, D. 2007, \mnras, 381, 1197

\bibitem[{{Hirata} \& {Seljak}(2004)}]{2004Hirata}
{Hirata}, C.~M., \& {Seljak}, U. 2004, \prd, 70, 063526

\bibitem[{{Hung} \& {Ebeling}(2012)}]{2012Hung}
{Hung}, C.-L., \& {Ebeling}, H. 2012, \mnras, 421, 3229

\bibitem[{{Joachimi} {et~al.}(2011){Joachimi}, {Mandelbaum}, {Abdalla}, \&
  {Bridle}}]{2011Joachimi}
{Joachimi}, B., {Mandelbaum}, R., {Abdalla}, F.~B., \& {Bridle}, S.~L. 2011,
  \aap, 527, A26

\bibitem[{{Kauffmann} {et~al.}(2003){Kauffmann}, {Heckman}, {White}, {Charlot},
  {Tremonti}, {Brinchmann}, {Bruzual}, {Peng}, {Seibert}, {Bernardi},
  {Blanton}, {Brinkmann}, {Castander}, {Cs{\'a}bai}, {Fukugita}, {Ivezic},
  {Munn}, {Nichol}, {Padmanabhan}, {Thakar}, {Weinberg}, \&
  {York}}]{2003bKauffmann}
{Kauffmann}, G., {Heckman}, T.~M., {White}, S.~D.~M., {Charlot}, S.,
  {Tremonti}, C., {Brinchmann}, J., {Bruzual}, G., {Peng}, E.~W., {Seibert},
  M., {Bernardi}, M., {Blanton}, M., {Brinkmann}, J., {Castander}, F.,
  {Cs{\'a}bai}, I., {Fukugita}, M., {Ivezic}, Z., {Munn}, J.~A., {Nichol},
  R.~C., {Padmanabhan}, N., {Thakar}, A.~R., {Weinberg}, D.~H., \& {York}, D.
  2003, \mnras, 341, 33

\bibitem[{{Komatsu} {et~al.}(2010){Komatsu}, {Smith}, {Dunkley}, {Bennett},
  {Gold}, {Hinshaw}, {Jarosik}, {Larson}, {Nolta}, {Page}, {Spergel},
  {Halpern}, {Hill}, {Kogut}, {Limon}, {Meyer}, {Odegard}, {Tucker}, {Weiland},
  {Wollack}, \& {Wright}}]{2010Komatsu}
{Komatsu}, E., {Smith}, K.~M., {Dunkley}, J., {Bennett}, C.~L., {Gold}, B.,
  {Hinshaw}, G., {Jarosik}, N., {Larson}, D., {Nolta}, M.~R., {Page}, L.,
  {Spergel}, D.~N., {Halpern}, M., {Hill}, R.~S., {Kogut}, A., {Limon}, M.,
  {Meyer}, S.~S., {Odegard}, N., {Tucker}, G.~S., {Weiland}, J.~L., {Wollack},
  E., \& {Wright}, E.~L. 2010, ArXiv e-prints

\bibitem[{{Li} {et~al.}(2013){Li}, {Jing}, {Faltenbacher}, \& {Wang}}]{2013Li}
{Li}, C., {Jing}, Y.~P., {Faltenbacher}, A., \& {Wang}, J. 2013, \apjl, 770,
  L12

\bibitem[{{Libeskind} {et~al.}(2012){Libeskind}, {Hoffman}, {Knebe},
  {Steinmetz}, {Gottl{\"o}ber}, {Metuki}, \& {Yepes}}]{2012Libeskind}
{Libeskind}, N.~I., {Hoffman}, Y., {Knebe}, A., {Steinmetz}, M.,
  {Gottl{\"o}ber}, S., {Metuki}, O., \& {Yepes}, G. 2012, \mnras, 421, L137

\bibitem[{{Mandelbaum} {et~al.}(2006){Mandelbaum}, {Seljak}, {Kauffmann},
  {Hirata}, \& {Brinkmann}}]{2006Mandelbaum}
{Mandelbaum}, R., {Seljak}, U., {Kauffmann}, G., {Hirata}, C.~M., \&
  {Brinkmann}, J. 2006, \mnras, 368, 715

\bibitem[{{Peebles}(1969)}]{1969Peebles}
{Peebles}, P.~J.~E. 1969, \apj, 155, 393

\bibitem[{{Pereira} \& {Kuhn}(2005)}]{2005Pereira}
{Pereira}, M.~J., \& {Kuhn}, J.~R. 2005, \apjl, 627, L21

\bibitem[{{Shen} {et~al.}(2003){Shen}, {Mo}, {White}, {Blanton}, {Kauffmann},
  {Voges}, {Brinkmann}, \& {Csabai}}]{2003Shen}
{Shen}, S., {Mo}, H.~J., {White}, S.~D.~M., {Blanton}, M.~R., {Kauffmann}, G.,
  {Voges}, W., {Brinkmann}, J., \& {Csabai}, I. 2003, \mnras, 343, 978

\bibitem[{{Tegmark} {et~al.}(2004){Tegmark}, {Blanton}, {Strauss}, {Hoyle},
  {Schlegel}, {Scoccimarro}, {Vogeley}, {Weinberg}, {Zehavi}, {Berlind},
  {Budavari}, {Connolly}, {Eisenstein}, {Finkbeiner}, {Frieman}, {Gunn},
  {Hamilton}, {Hui}, {Jain}, {Johnston}, {Kent}, {Lin}, {Nakajima}, {Nichol},
  {Ostriker}, {Pope}, {Scranton}, {Seljak}, {Sheth}, {Stebbins}, {Szalay},
  {Szapudi}, {Verde}, {Xu}, {Annis}, {Bahcall}, {Brinkmann}, {Burles},
  {Castander}, {Csabai}, {Loveday}, {Doi}, {Fukugita}, {Gott}, {Hennessy},
  {Hogg}, {Ivezi{\'c}}, {Knapp}, {Lamb}, {Lee}, {Lupton}, {McKay}, {Kunszt},
  {Munn}, {O'Connell}, {Peoples}, {Pier}, {Richmond}, {Rockosi}, {Schneider},
  {Stoughton}, {Tucker}, {Vanden Berk}, {Yanny}, {York}, \& {SDSS
  Collaboration}}]{2004Tegmark}
{Tegmark}, M., {Blanton}, M.~R., {Strauss}, M.~A., {Hoyle}, F., {Schlegel}, D.,
  {Scoccimarro}, R., {Vogeley}, M.~S., {Weinberg}, D.~H., {Zehavi}, I.,
  {Berlind}, A., {Budavari}, T., {Connolly}, A., {Eisenstein}, D.~J.,
  {Finkbeiner}, D., {Frieman}, J.~A., {Gunn}, J.~E., {Hamilton}, A.~J.~S.,
  {Hui}, L., {Jain}, B., {Johnston}, D., {Kent}, S., {Lin}, H., {Nakajima}, R.,
  {Nichol}, R.~C., {Ostriker}, J.~P., {Pope}, A., {Scranton}, R., {Seljak}, U.,
  {Sheth}, R.~K., {Stebbins}, A., {Szalay}, A.~S., {Szapudi}, I., {Verde}, L.,
  {Xu}, Y., {Annis}, J., {Bahcall}, N.~A., {Brinkmann}, J., {Burles}, S.,
  {Castander}, F.~J., {Csabai}, I., {Loveday}, J., {Doi}, M., {Fukugita}, M.,
  {Gott}, III, J.~R., {Hennessy}, G., {Hogg}, D.~W., {Ivezi{\'c}}, {\v Z}.,
  {Knapp}, G.~R., {Lamb}, D.~Q., {Lee}, B.~C., {Lupton}, R.~H., {McKay}, T.~A.,
  {Kunszt}, P., {Munn}, J.~A., {O'Connell}, L., {Peoples}, J., {Pier}, J.~R.,
  {Richmond}, M., {Rockosi}, C., {Schneider}, D.~P., {Stoughton}, C., {Tucker},
  D.~L., {Vanden Berk}, D.~E., {Yanny}, B., {York}, D.~G., \& {SDSS
  Collaboration}. 2004, \apj, 606, 702

\bibitem[{{The Enzo Collaboration} {et~al.}(2013){The Enzo Collaboration},
  {Bryan}, {Norman}, {O'Shea}, {Abel}, {Wise}, {Turk}, {Reynolds}, {Collins},
  {Wang}, {Skillman}, {Smith}, {Harkness}, {Bordner}, {Kim}, {Kuhlen}, {Xu},
  {Goldbaum}, {Hummels}, {Kritsuk}, {Tasker}, {Skory}, {Simpson}, {Hahn},
  {Oishi}, {So}, {Zhao}, {Cen}, \& {Li}}]{2013Enzo}
{The Enzo Collaboration}, {Bryan}, G.~L., {Norman}, M.~L., {O'Shea}, B.~W.,
  {Abel}, T., {Wise}, J.~H., {Turk}, M.~J., {Reynolds}, D.~R., {Collins},
  D.~C., {Wang}, P., {Skillman}, S.~W., {Smith}, B., {Harkness}, R.~P.,
  {Bordner}, J., {Kim}, J.-h., {Kuhlen}, M., {Xu}, H., {Goldbaum}, N.,
  {Hummels}, C., {Kritsuk}, A.~G., {Tasker}, E., {Skory}, S., {Simpson}, C.~M.,
  {Hahn}, O., {Oishi}, J.~S., {So}, G.~C., {Zhao}, F., {Cen}, R., \& {Li}, Y.
  2013, ArXiv e-prints

\bibitem[{{Torlina} {et~al.}(2007){Torlina}, {De Propris}, \&
  {West}}]{2007Torlina}
{Torlina}, L., {De Propris}, R., \& {West}, M.~J. 2007, \apjl, 660, L97

\bibitem[{{Trujillo} {et~al.}(2006){Trujillo}, {F{\"o}rster Schreiber},
  {Rudnick}, {Barden}, {Franx}, {Rix}, {Caldwell}, {McIntosh}, {Toft},
  {H{\"a}ussler}, {Zirm}, {van Dokkum}, {Labb{\'e}}, {Moorwood},
  {R{\"o}ttgering}, {van der Wel}, {van der Werf}, \& {van
  Starkenburg}}]{2006Trujillo}
{Trujillo}, I., {F{\"o}rster Schreiber}, N.~M., {Rudnick}, G., {Barden}, M.,
  {Franx}, M., {Rix}, H., {Caldwell}, J.~A.~R., {McIntosh}, D.~H., {Toft}, S.,
  {H{\"a}ussler}, B., {Zirm}, A., {van Dokkum}, P.~G., {Labb{\'e}}, I.,
  {Moorwood}, A., {R{\"o}ttgering}, H., {van der Wel}, A., {van der Werf}, P.,
  \& {van Starkenburg}, L. 2006, \apj, 650, 18

\bibitem[{{Turk} {et~al.}(2011){Turk}, {Smith}, {Oishi}, {Skory}, {Skillman},
  {Abel}, \& {Norman}}]{2011Turk}
{Turk}, M.~J., {Smith}, B.~D., {Oishi}, J.~S., {Skory}, S., {Skillman}, S.~W.,
  {Abel}, T., \& {Norman}, M.~L. 2011, \apjs, 192, 9

\bibitem[{{Vitvitska} {et~al.}(2002){Vitvitska}, {Klypin}, {Kravtsov},
  {Wechsler}, {Primack}, \& {Bullock}}]{2002Vitvitska}
{Vitvitska}, M., {Klypin}, A.~A., {Kravtsov}, A.~V., {Wechsler}, R.~H.,
  {Primack}, J.~R., \& {Bullock}, J.~S. 2002, \apj, 581, 799

\bibitem[{{Zhang} {et~al.}(2013){Zhang}, {Yang}, {Wang}, {Wang}, {Mo}, \& {van
  den Bosch}}]{2013Zhang}
{Zhang}, Y., {Yang}, X., {Wang}, H., {Wang}, L., {Mo}, H.~J., \& {van den
  Bosch}, F.~C. 2013, \apj, 779, 160

\end{thebibliography}

\end{document}